\documentclass[conference]{IEEEtran}

\IEEEoverridecommandlockouts
\usepackage{cite}
\usepackage{overpic}
\usepackage{amsmath,amssymb,amsfonts}
\usepackage{graphicx}
\usepackage{textcomp}
\usepackage{xcolor}
\usepackage{float}
\usepackage{amsthm}
\usepackage{graphicx}
\usepackage{epstopdf}
\usepackage{amsmath,bm}
\usepackage{amsfonts}
\usepackage{amssymb}
\usepackage{color}
\usepackage{multirow}
\usepackage{multicol}
\usepackage{soul,xcolor}
\usepackage{algorithm}
\usepackage{algpseudocode}%

\theoremstyle{plain}

\newcommand{\vect}[1]{\mathbf{#1}}

\def\diag{\mathrm{diag}}

\def\kron{\otimes}

\def\Htran{\mbox{\tiny $\mathrm{H}$}}
\def\Ttran{\mbox{\tiny $\mathrm{T}$}}
\def\imagunit{\mathsf{j}} 

\newcommand{\argmax}[1]{{\underset{{#1}}{\mathrm{arg\,max}}}}

\begin{document}

\title{Predictive Beamforming with Distributed MIMO}
\author{\IEEEauthorblockN{
Hasret Taha Akçalı\IEEEauthorrefmark{1}\IEEEauthorrefmark{2}, \"Ozlem Tu\u{g}fe Demir\IEEEauthorrefmark{2}, Tolga Girici\IEEEauthorrefmark{2}, and Emil Björnson\IEEEauthorrefmark{3}
}
\IEEEauthorblockA{\IEEEauthorrefmark{1}Communications and Information Technologies Business Sector, ASELSAN A.Ş., Ankara, Türkiye}
\IEEEauthorblockA{\IEEEauthorrefmark{2}Department of Electrical-Electronics Engineering, TOBB ETU, Ankara, Türkiye} (takcali@aselsan.com.tr) (\{hakcali, ozlemtugfedemir, tgirici\}@etu.edu.tr)
\IEEEauthorblockA{\IEEEauthorrefmark{3}Department of Computer Science, KTH Royal Institute of Technology, Stockholm, Sweden (emilbjo@kth.se)}
}

\maketitle

\begin{abstract}
In vehicle-to-everything (V2X) applications, roadside units (RSUs) can be tasked with both sensing and communication functions to enable sensing-assisted communications. Recent studies have demonstrated that distance, angle, and velocity information obtained through sensing can be leveraged to reduce the overhead associated with communication beam tracking. In this work, we extend this concept to scenarios involving multiple distributed RSUs and distributed MIMO (multiple-input multiple-output) systems. We derive the state evolution model, formulate the extended Kalman-filter equations, and implement predictive beamforming for distributed MIMO. Simulation results indicate that, when compared with a co-located massive MIMO antenna array, distributed antennas lead to more uniform and robust sensing performance, coverage, and data rates, while the vehicular user is in motion.
\end{abstract}


\section{Introduction}

Millimeter-wave communications (mmWave) plays a crucial role in next-generation wireless connectivity, offering high bandwidth capabilities \cite{7999294}. However, mmWave signals face significant challenges, especially signal scattering and blockage. These issues can be mitigated by employing massive MIMO (multiple-input multiple-output) arrays, which enable the formation of narrow, high-gain beams. However, the training and tracking process for a massive number of antennas can introduce considerable overhead, which presents a critical challenge to be addressed \cite{8599167}.

In addition to their communication benefits, the high carrier frequency and large bandwidth of 5G mmWave signals make them well-suited for radar sensing applications. With high delay and Doppler resolutions, as well as centimeter-level localization accuracy, 5G mmWave signals can be applied in various domains, including automotive and vehicle-to-everything (V2X) communications \cite{9162963}. V2X communication in 5G and beyond, is a scenario that requires low-latency and high throughput. Especially in autonomous driving, sensing capability is essential for obstacle detection and accurate localization with centimeter accuracy.  \cite{Liu2020RadarAssisted}  Traditionally, radar and communication functions have been carried out using separate hardware and in different frequency bands. However, as the number of radar and communication applications grows, these frequency bands converge, making efficient spectrum utilization increasingly critical. Integrated sensing and communications (ISAC) has recently emerged as a prominent research area focusing on utilizing the same resources for both sensing and communication \cite{ITU-IMT2030}. ISAC not only optimizes spectrum usage but also reduces costs, size, and weight by allowing the joint operation of sensing and communication on the same hardware \cite{ITU-IMT2030}.

It has been recently proven that information obtained by on-board sensors such as cameras \cite{zhang2022camera}, LIDAR \cite{jiang2022lidar}, inertial sensors mounted on antennas, GPS measurements \cite{ciaramitaro2023signalling} and vehicle speed information \cite{xu2021predictive}   can be useful in decreasing the beam tracking overhead in V2X communications. Recently, the concept of radar-assisted predictive beamforming has been demonstrated in a point-to-point link with a single road-side unit (RSU) \cite{Liu2020RadarAssisted, liu2022learning}. In this approach, estimated distance, angle, and reflection coefficient obtained through radar sensing are used to predict future angle values, allowing for the design of optimal precoding and combining vectors at both the RSU and the vehicle. This work is later extended in \cite{du2022integrated} for extended targets, rather than point targets.  While this method achieves meaningful improvement in data rates, variations can occur as the vehicular user moves. The radar accuracy depends on the distance (range) and as the vehicle moves away from the RSU, the radar information loses accuracy, affecting communication performance. Therefore, it is a promising solution to utilize multiple RSU's to improve robustness in performance. 

For the first time in the literature, we introduce predictive beamforming with \emph{distributed MIMO} to address these data rate variations. This is inspired by the distributed MIMO-ISAC systems that provide superior target localization and detection performance, as well as enhance the spectral efficiency (SE) \cite{10236534}. In the proposed distributed MIMO-ISAC network, a group of RSUs equipped with joint radar and communication transceivers are strategically distributed across the network's coverage area. This architecture aims to not only enhance parameter estimation accuracy but also boost SE for vehicles when compared to a single massive MIMO transceiver.

In this work, we derive state prediction equations of the the extended Kalman filter (EKF) to predict and track the distance and velocity of a vehicle, as the problem is inherently nonlinear, making EKF more suitable than the standard Kalman filter. This sensing information is then utilized for precise beamforming, enhancing communication performance. We evaluate the performance and robustness of the proposed approach through a comprehensive simulation study.

\section{System Model} \label{System Model Section}
\begin{figure}[h]
    \centering
    \includegraphics[width=\columnwidth]{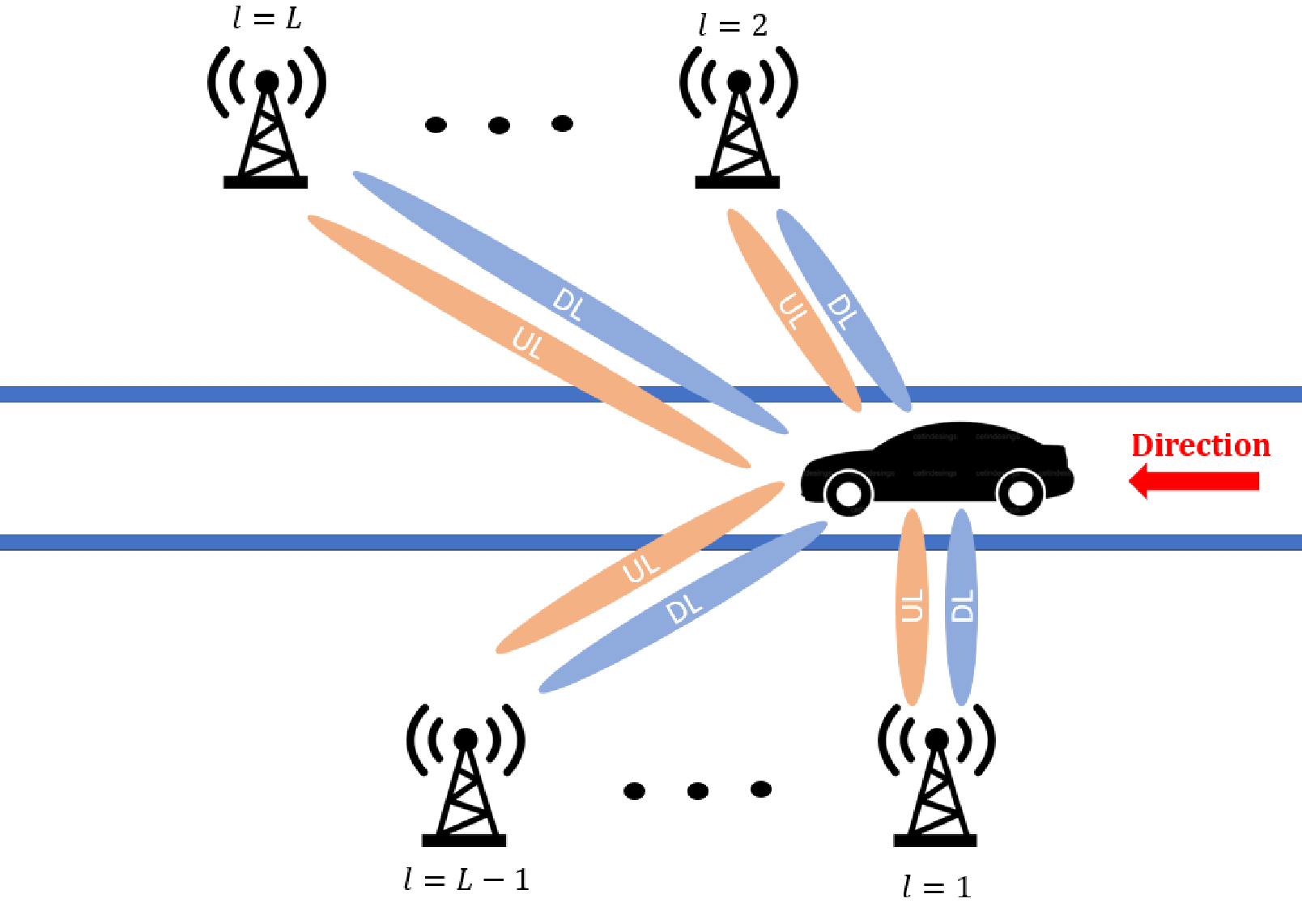}
    \caption{Illustration of the system model, where the RSUs are connected to a CPU (not shown in the figure). }

    \label{System Model}
\end{figure}

The system considered is shown in Fig.~\ref{System Model} and consists of $L$ RSUs and a vehicle, each of which is equipped with a horizontal uniform linear array (ULA) and positioned parallel to the road. Each RSU is equipped with $N_t$ transmit and $N_r$ receive antennas, while the vehicle is equipped with $M$ transmit and receive antennas. The vehicle moves from the right side to the left side at a constant velocity. The channel between the RSUs and the vehicle is assumed to be line-of-sight (LoS) without any multipath fading. For simplicity in analysis, we also assume that each RSU transmits using orthogonal waveforms (as will be defined in Section~\ref{RSU Radar Signal Model}), thereby eliminating interference concerns. Although not shown in Fig. \ref{System Model}, we assume that each RSU is connected via fronthaul to a central processing unit (CPU), which coordinates the transmission of RSUs. 

In 5G, beam alignment is achieved through beam training and beam tracking methods, which are based on reference signals and feedback signals. In this article, we explore the prediction of the beam assisted by sensing, a novel method that can be applied in 6G for ISAC \cite{Liu2020RadarAssisted, liu2022learning}. 
\begin{figure*}[h]
    \centering
    \includegraphics[width=1.5\columnwidth]{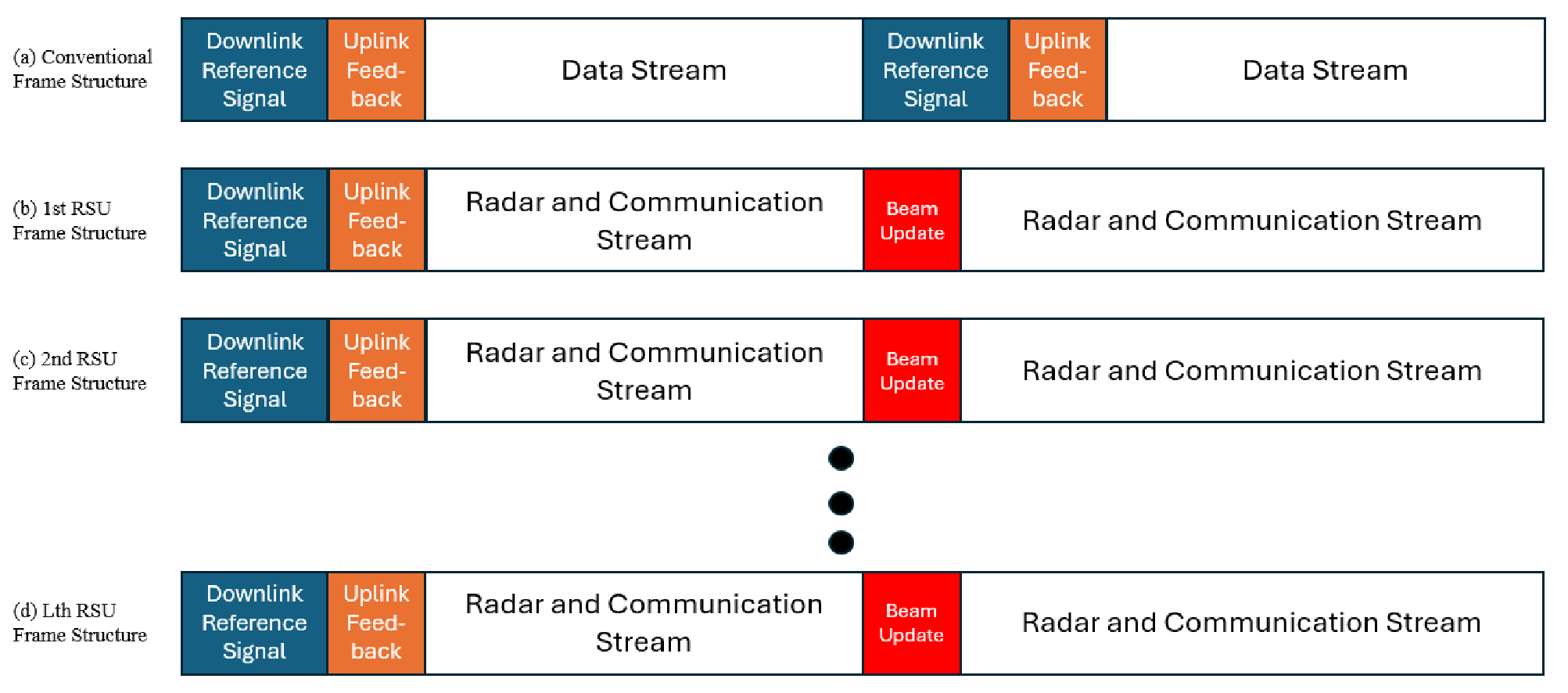}
    \caption{Frame structure of the proposed approach. } 
    \label{Frame Structure}
\end{figure*}

As shown in Fig. \ref{Frame Structure}, only the initial beams are detected using the conventional method. Beam prediction parameters are analyzed using radar and communication streams at the CPU. Additionally, vehicle beams are periodically updated based on beam information received from RSUs. A comparison between conventional beam tracking methods and the proposed approach is not included in this paper, as it has been previously addressed in \cite{Liu2020RadarAssisted}. Instead, this work examines the predictive beamforming approach in the context of distributed MIMO for the first time.   
\subsection{State Evolution Model} 
\begin{figure}[h]
    \centering
    \includegraphics[width=\columnwidth]{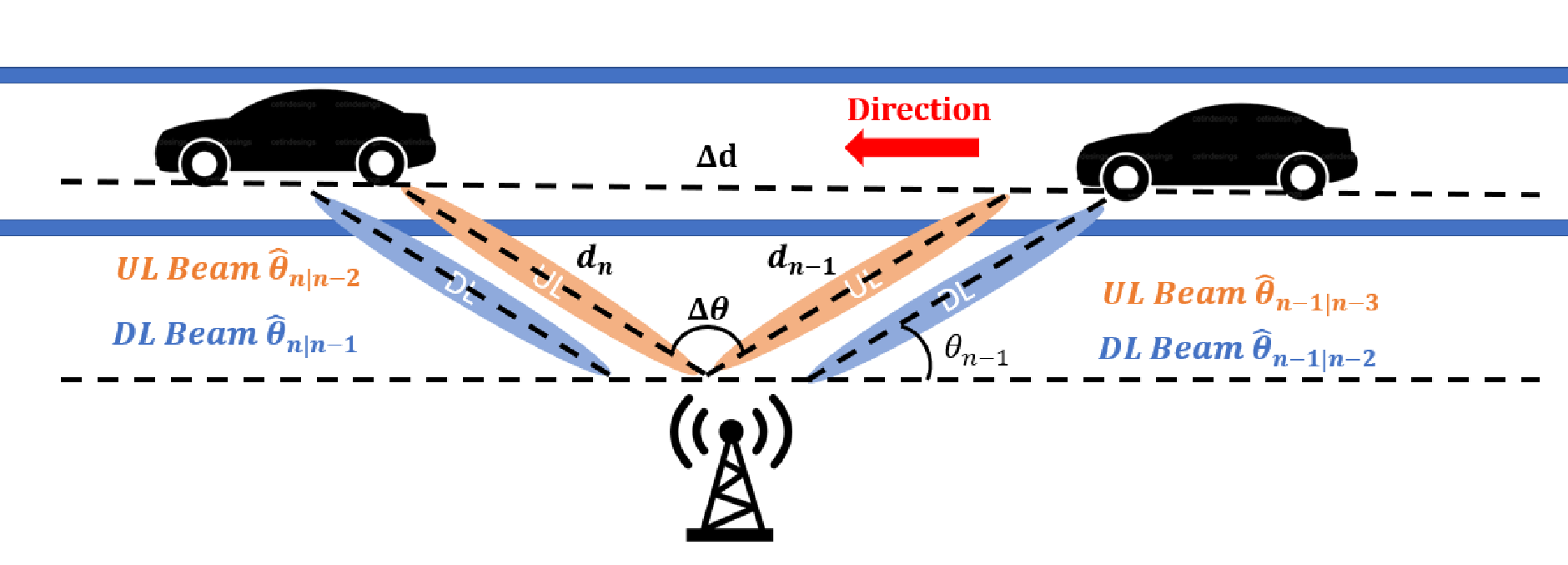}
    \caption{V2X state evolution model. } 
    \label{V2I State Evaluation Model}
\end{figure}

Using kinematic equations, each RSU determines the variations in the angle, distance, reflection coefficient, and velocity of the vehicle. At the $n^{th}$ epoch, the distance, angle, and reflection coefficient of the vehicle relative to the $l^{th}$ RSU and the velocity of the vehicle are evaluated. The reflection coefficient covers both the radar cross section of the vehicle and the loss along the signal path. The parameters $d_{l,n}$, $\theta_{l,n}$, and $\beta_{l,n}$ represent the distance, angle, and reflection coefficient of the vehicle relative to the $l^{th}$ RSU, respectively. The vehicle's velocity at the $n^{th}$ epoch is indicated by $v_{n}$. In the following section, we derive the state evolution model. Based on the 2D-geometry shown in Fig.~\ref{V2I State Evaluation Model}, we have
\begin{align}
&  d_{l,n}^2 = d_{l,n-1}^2 + \Delta d^2 - 2d_{l,n-1}  \Delta d \cos (\theta_{l,n-1}), \label{eq:state1}\\
&  \frac{\Delta d}{  \sin( \Delta\theta)} = \frac{d_{l,n}}{ \sin (\theta_{l,n-1})},\label{eq:state2}
\end{align}
where $\Delta d =  v_{n-1} \Delta T $, $\Delta\theta = \theta_{l,n} - \theta_{l,n-1}$. The indices $n$ and $n-1$ represent the current epoch and the previous epoch, respectively. $\Delta T$ is the time slot (epoch) duration.  From the bi-static radar range equation, we have the reflection coefficient from RSU $l$ to RSU $m$ in the time slot $n$ as
\begin{align}
    \beta_{l,m,n} = \frac{\lambda}{(4\pi)^{1.5}}\frac{\varepsilon_{l,m}}{d_{l,n}d_{m,n}} \label{eq:state3}
\end{align}
where $\varepsilon_{l,m}$ is the complex radar cross section corresponding the two-way path between the $l$th RSU and $m$th RSU, which is assumed to be constant throughout the observation interval according to the Swerling I model \cite{richards2010principles}.

It is challenging
to analyze the evolution model directly given the highly non-linear nature of \eqref{eq:state1} and \eqref{eq:state2}. Thus, by the help of \cite{Liu2020RadarAssisted}, we get the approximations
\begin{align}
 &   \theta_{l,n} \approx \theta_{l,n-1} + d_{l,n-1}^{-1} v_{n-1} \Delta T \sin (\theta_{l,n-1}), \\
&    d_{l,n}      \approx d_{l,n-1} - v_{n-1}  \Delta T \cos( \theta_{l,n-1}) , \\
 &   v_{n}       \approx v_{n-1}.
\end{align}To approximate the state evolution of $\beta_{l,m,n}$, we note that
\begin{align}
    \beta_{l,m,n}& = \beta_{l,m,n-1}\frac{d_{l,n-1}d_{m,n-1}}{d_{l,n}d_{m,n}} \\
    &\approx \beta_{l,m,n-1}\left(1+\frac{\Delta d \cos(\theta_{l,n-1})}{d_{l,n}}\right)\nonumber\\
    &\quad \times\left(1+\frac{\Delta d \cos(\theta_{m,n-1})}{d_{m,n}}\right) \nonumber\\
    &\approx \beta_{l,m,n-1}\left(1+\frac{\Delta d \cos(\theta_{l,n-1})}{d_{l,n-1}}+\frac{\Delta d \cos(\theta_{m,n-1})}{d_{m,n-1}}\right). \nonumber
\end{align}
Finally, the state evolution model becomes as shown in \eqref{eq:state4} at the top of the page after the next. Here, $w_{v,n}$, $w_{\theta,l,n}$, $w_{d,l,n}$, and $w_{\beta,l,m,n}$ represent the corresponding noises, which are assumed to be zero-mean Gaussian distributed with variances of $\sigma_v^2$, $\sigma_{\theta,l}^2$, $\sigma_{d,l}^2$, and $\sigma_{\beta,l,m}^2$, respectively.  Note that these noises stand only for representing approximation and other systematic errors and the respective statistics are utilized in the state prediction of extended Kalman filtering.

\begin{figure*}

\begin{align}
    \left\{
    \begin{array}{ll}  v_{n}        = v_{n-1} + w_{v,n}, \\
    \theta_{l,n} = \theta_{l,n-1} + d_{l,n-1}^{-1} v_{n-1} \Delta T \sin (\theta_{l,n-1}) + w_{\theta,l,n}, \quad l=1,\ldots,L \\
    d_{l,n}      = d_{l,n-1} - v_{n-1}  \Delta T \cos( \theta_{l,n-1}) + w_{d,l,n},  \quad l=1,\ldots,L  \\
    \beta_{l,m,n}  = \beta_{l,m,n-1} \left(1 + d_{l,n-1}^{-1} v_{n-1} \Delta T \cos( \theta_{l,n-1})+d_{m,n-1}^{-1} v_{n-1} \Delta T \cos( \theta_{m,n-1})\right) + w_{\beta,l,m,n}, ~ l,m=1,\ldots, L.
    \end{array}
    \right.
    \label{eq:state4}
\end{align}
\hrulefill
\end{figure*}

\subsection{Signal Model}
In the following subsections, we sequentially discuss the radar signal and measurement model, followed by the communication model. We assume that the initial estimates of the parameters of interest have already been obtained through conventional radar signal processing tasks before the communication begins.

\subsubsection{Radar Signal Model} \label{RSU Radar Signal Model}
The RSUs send orthogonal waveforms $s_{l,n}(t)$, i.e., $\int_{0}^{L T} s_{l,n}(t)s_{m,n}(t)dt = 0$ for $m\neq l$, where $T$ is the symbol period if there was only one RSU with a certain bandwidth. The orthogonality is ensured through orthogonal waveforms of $LT$ length, which have the same bandwidth as in the case of one RSU. RSU $l$ multiplies $s_{l,n}(t)$ with the unit-norm precoding vector $\vect{f}_{l,n}$ and transmits the signal $\sqrt{p_n}\vect{f}_{l,n} s_{l,n}(t)$, where $p_n$ is the transmit power of each RSU at the $n^{th}$ epoch.

\subsubsection{Radar Measurement Model}

The reflected echo from the vehicle at RSU $l$ is \begin{align}
   \vect{r}_{m,n}(t) = \sqrt{p_n}\sum_{l=1}^L&\beta_{l,m,n}e^ {\imagunit2\pi\mu_{l,m,n}t}\vect{b}(\theta_{m,n}) \vect{a}^{\Htran}(\theta_{l,n}) \nonumber\\
&\times\vect{f}_{l,n} s_{l,n}(t- \tau_{l,m,n})+\vect{z}_{l,n}(t),
    \label{eq:state5}
\end{align}
where the measurement noise is denoted by $\vect{z}_{l,n}(t)$. The $\vect{a}(\theta)$ and $\vect{b}(\theta)$ are the transmit and receive steering vector of the antenna array of the RSU, which are expressed as
\begin{align}
& \vect{a}(\theta) = \begin{bmatrix}  1 & e^{-\imagunit \pi \cos ({\theta})} & \cdots & e^{-\imagunit \pi(N_t-1) \cos ({\theta})} \end{bmatrix}^{\Ttran}, \\
& \vect{b}(\theta) = \begin{bmatrix}  1 & e^{-\imagunit \pi \cos ({\theta})} & \cdots & e^{-\imagunit \pi(N_r-1) \cos ({\theta})} \end{bmatrix}^{\Ttran}, 
\end{align}
where ULAs with half-wavelength antenna spacing are assumed.  Moreover, the estimates of pairwise delays $\tau_{l,m,n}$ and pairwise Doppler shifts are mathematically given as
\begin{align}
    &\{\hat{\tau}_{l,m,n},\hat{\mu}_{l,m,n}\} = \nonumber\\& \argmax{\tau,\mu} \left| \int_0^{\Delta T } \vect{r}_{l,n}(t)s_{m,n}^*(t-\tau)e^{-\imagunit2\pi\mu t}d t\right|^2
\end{align}
for $m=1,\ldots,L$. Substituting the above estimates into \eqref{eq:state5}, we obtain 
\begin{align}
&\tilde{\vect{r}}_{l,r,n}=\int_{0}^{\Delta T} \vect{r}_{l,n}(t)s_{r,n}^*(t - \hat{\tau}_{l,r,n})e^{-\imagunit 2\pi\hat{\mu}_{l,r,n}t} dt \nonumber\\
&\approx \sqrt{p_{n}G}\beta_{l,r,n}\vect{b}(\theta_{l,n})\vect{a}^{\Htran}(\theta_{r,n})\vect{f}_{r,n}+ \vect{z}_{\vect{r},l,r,n},
\end{align}
where $G$ is the matched filtering gain (pulse compression gain) and we assumed the small contributions from the waveforms other than $s_{r,n}(t)$ are added into the noise term. The noise vector is distributed as $\vect{z}_{\vect{r},l,r,n}\sim\mathcal{N}_{\mathbb{C}}(\vect{0},\sigma^2_1\vect{I}_{N_r})$
The measurements related to delays and Doppler shifts are given as \cite{du2014explicit}
\begin{align}
& \tau_{l,m,n} = \frac{d_{l,n}+ d_{m,n}}{c}+z_{\tau,l,m,n}, \\
& \mu_{l,m,n} = \frac{v_n(\cos\theta_{l,n}+\cos\theta_{m,n})f_c}{c}+ z_{\mu,l,m,n},
\end{align}
where $f_c$ and $c$ represent the carrier frequency and the speed of light, respectively. The $z_{\tau,l,m,n}$ and $z_{\mu,l,m,n}$ are the respective Gaussian measurement noises with variances $\sigma_{\tau,l,m,n}^2$ and $\sigma_{\mu,l,m,n}^2$, respectively. 

The precoding vectors $\vect{f}_{l,n}$  are designed based on prediction of angles at the RSU side given as $\vect{f}_{l,n} = \frac{\vect{a}\left(\hat{\theta}_{l,n|n-1}\right)}{\sqrt{N_t}}$, where $\hat{\theta}_{l,n|n-1}$ is the one-step predicted angle for the $l^{th}$ RSU at the $n$th epoch. The radar measurement model for RSU $l$ is expressed as shown in \eqref{eq:state13} at the top of the page after the next. Note that variances of measurement noise are inversely proportional to the receive signal-to-noise ratio (SNR) in (\ref{eq:state5}). Based on (\ref{eq:state5}), the measurement variances $\sigma_{\tau,l,m,n}^2$ and $\sigma_{\mu,l,m,n}^2$ expressed as
\begin{align}
& \sigma^2_{\tau,l,m,n} =  \frac{\sigma^2_{2}}{GN_r|\beta_{l,m,n}|^2|\delta_{m,n}|^2p_n} \\
&  \sigma^2_{\mu,l,m,n}=  \frac{\sigma^2_3}{GN_r|\beta_{l,m,n}|^2|\delta_{m,n}|^2p_n}
\end{align}
where $\delta_{m,n} = \vect{a}^{\Htran}(\theta_{m,n})\vect{a}(\hat{\theta}_{m,n|n-1})/\sqrt{N_t} $.

\begin{figure*}
\begin{align}
    \left\{
    \begin{array}{ll}
    &     \tilde{\vect{r}}_{l,m,n}(t) = \sqrt{p_nG/N_t} \beta_{l,m,n}\vect{b}(\theta_{l,n})\vect{a}^{\Htran}(\theta_{m,n})\vect{a}\left(\hat{\theta}_{m,n|n-1}\right)+\vect{z}_{\vect{r},l,m,n}, \quad l,m=1,\ldots,L \\
    & \tau_{l,m,n} = \frac{d_{l,n}+ d_{m,n}}{c}+z_{\tau,l,m,n}, \quad l,m=1,\ldots,L \\
    & \mu_{l,m,n} = \frac{v_n(\cos\theta_{l,n}+\cos\theta_{m,n})f_c}{c}+ z_{\mu,l,m,n}, \quad l,m=1,\ldots,L
    \end{array}
    \right.
    \label{eq:state13}
\end{align}
\hrulefill
\end{figure*}
\subsubsection{Communication Model}
At the $n^{th}$ epoch, using a filter matched to the $l^{th}$ RSU's waveform, the vehicle receives the signal from $l^{th}$ RSU by using a receive beamformer $\vect{w}_{l,n}$, yielding
\begin{align}
    \vect{c}_{l,n}(t) = & \sqrt{p_{n}L}\alpha_{l,n}e^{\imagunit2\pi\varrho_{l,n}t}\vect{w}_{l,n}^{\Htran}\vect{u}(\theta_{l,n})\vect{a}^{\Htran}(\theta_{l,n})\vect{f}_{l,n}s_{l,n} \nonumber\\&+ z_{c,l,n}(t), \quad l=1,\ldots,L,
\label{eq:state14}
\end{align}where $ z_{c,l,n}(t)$ is a zero-mean white Gaussian noise process with variance $\sigma_c^2$, $s_{l,n}$ is the unit-power  information symbol embedded into the waveform $s_{l,n}(t)$, $\alpha_{l,n}$ denotes $l^{th}$ RSU communication channel coefficient at $n^{th}$ epoch, $\vect{w}_{l,n}$ denotes the receive beamforming vector at $n^{th}$ epoch, and $\varrho_{l,n}=\frac{v_n\cos(\theta_{l,n})f_c}{c}$ is the Doppler frequency of vehicle with respect to $l^{th}$ RSU at $n^{th}$ epoch. Moreover, $\vect{u}(\theta)$ is the transmit and receive steering vector of the antenna array of the vehicle, which are expressed as 
\begin{equation}
    \vect{u}(\theta) = [1,   e^{-\imagunit \pi \cos ({\theta})} ,  \cdots  , e^{-\imagunit \pi(M-1) \cos ({\theta})}]^{\Ttran}  
\end{equation} where the ULA has half-wavelength antenna spacing.

The parameter of the receive beamforming vector should be two-step prediction of the angle, since one-step prediction of the angle would be out-of-date. Thus, the receive beamforming vector at vehicle expressed as  $\vect{w}_{n} = \frac{\vect{u}\left(\hat{\theta}_{n|n-2}\right)}{\sqrt{N_r}}$,  where the parameter $\hat{\theta}_{n|n-2}$ and its calculation is expressed in Section \ref{Proposed Approach}. At each epoch, each RSU generates only one beam, as there is a single vehicle to track. Therefore, all power is concentrated in the single beam of each RSU. Since the vehicle is aware of its own velocity and can estimate $\theta_{l,n}$ by receiving the angle prediction from the RSU, the Doppler shift at the receiver side can be compensated \cite{Liu2020RadarAssisted}. The resulting receive SNR at the vehicle is obtained as
\begin{align}
 & \mathrm{SNR}_{l,n} = \frac{p_{n}L \left|\alpha_{l,n}\vect{w}_{l,n}^{\Htran}\vect{u}(\theta_{l,n})\vect{a}^{\Htran}(\theta_{l,n})\vect{f}_{l,n}\right|^2}{\sigma_c^2}  \\
 & =  \frac{p_{n} L\left|\alpha_{l,n}\vect{u}^{\Htran}\left(\hat{\theta}_{l,n|n-2}\right)\vect{u}(\theta_{l,n})\vect{a}^{\Htran}(\theta_{l,n})\vect{a}\left(\hat{\theta}_{l,n|n-1}\right)\right|^2}{N_tN_r\sigma_c^2} \nonumber
\end{align}
where $\alpha_{l,n}$ is LoS channel coefficient with the corresponding path loss given as $|\alpha_{l,n}|^2 = \frac{\Upsilon}{d_{l,n}^{\alpha}} $
where $\Upsilon$ is the channel power gain at the reference distance $d = 1$\,m and $\alpha$ is the path loss exponent, which are known at the RSU. On the other hand, the RSUs send orthogonal waveforms, which is mentioned in Section \ref{RSU Radar Signal Model}. To sum up, the SE of the vehicle is calculated as
\begin{align}
   \mathrm{SE} =\frac{1}{L} \sum_{l=1}^L\log_2(1+\mathrm{SNR}_{l,n}) 
    \label{eq:state33}
\end{align}where the pre-log factor $1/L$ is added since the symbol period is increased by $L$ to assign orthogonal waveforms with the same bandwidth. Downlink pilots are assumed to be transmitted in each block, enabling the vehicle to estimate the channel and achieve the aforementioned SE. Radar sensing is also used to estimate the propagation delays from different RSUs. Therefore this knowledge can be exploited by the controller to adjust phase shifts of the signals from different RSUs for coherent combining at the communications receiver. 
\section{Proposed Approach} \label{Proposed Approach}

In this section, we will formulate the EKF equations for predictive beamforming in the presence of distributed MIMO antennas.  The Kalman filter is an important mathematical tool, primarily used for the purpose of estimating the state of a linear dynamic system from a series of noisy measurements. In our case, nonlinearity in our state equations (like \eqref{eq:state1}-\eqref{eq:state2}) and nonlinear measurement or observation equations can be eliminated with the help of the EKF approach. EKF approach performs local linearization for nonlinear models. In this paper, nonlinear models are expressed as
 \vspace{-0.1in}
\begin{align}
&     \text{State Evaluation Model: }  \vect{x}_n =\vect{g}(\vect{x}_{n-1}) + \vect{w}_n \\
&     \text{Measurement Model: } \vect{y}_n =\vect{h}(\vect{x}_n) + \vect{z}_n
\end{align}where $\vect{g}(\cdot)$ and $\vect{h}(\cdot)$ are defined in (\ref{eq:state4}) and (\ref{eq:state13}), respectively. Here, 
$\vect{x}_n = [v_n, \boldsymbol{\theta}_n^{\Ttran}, \boldsymbol{d}_n^{\Ttran}, \boldsymbol{\beta}_n^{\Ttran}]^{\Ttran}$, $\vect{y}_n = [\Tilde{\vect{r}}_n^{\Ttran}, \boldsymbol{\tau}_n^{\Ttran}, \boldsymbol{\mu}_n^{\Ttran}]^{\Ttran}$ represent the following state variables and measured signal vectors,  respectively: 
\begin{align}
&\boldsymbol{\theta}_n = \begin{bmatrix} \theta_{1,n}, \ \cdots, \ \theta_{L,n}\end{bmatrix}^{\Ttran} \\
& \boldsymbol{d}_n = \begin{bmatrix} d_{1,n}, \ \cdots, \ d_{L,n}\end{bmatrix}^{\Ttran} \\
& \boldsymbol{\beta}_n = \begin{bmatrix} \beta_{1,1,n}, \ \cdots, \ \beta_{1,L,n}, \ \cdots, \ \beta_{L,1,n}, \ \cdots, \ \beta_{L,L,n}\end{bmatrix}^{\Ttran} 
\end{align}
\begin{align}
& \Tilde{\boldsymbol{r}}_n = \begin{bmatrix} \Tilde{\boldsymbol{r}}_{1,1,n}^{\Ttran}, \ \cdots, \ \Tilde{\boldsymbol{r}}_{1,L,n}^{\Ttran}, \ \cdots, \ \Tilde{\boldsymbol{r}}_{L,1,n}^{\Ttran}, \ \cdots, \ \Tilde{\boldsymbol{r}}_{L,L,n}^{\Ttran}  \end{bmatrix}^{\Ttran} \\
& \boldsymbol{\tau}_n = \begin{bmatrix} \tau_{1,1,n}, \ \cdots, \ \tau_{1,L,n}, \ \cdots, \ \tau_{L,1,n}, \ \cdots, \ \tau_{L,L,n}\end{bmatrix}^{\Ttran} \\
& \boldsymbol{\mu}_n = \begin{bmatrix} \mu_{1,1,n}, \ \cdots, \ \mu_{1,L,n}, \ \cdots, \ \mu_{L,1,n}, \ \cdots, \ \mu_{L,L,n}\end{bmatrix}^{\Ttran}.
\end{align}
The related models are defined in  \eqref{eq:state4} and \eqref{eq:state13} with the state evaluation noise vector $\vect{w}_n= [w_{v,n}, \vect{w}_{\theta,n}^{\Ttran}, \vect{w}_{d,n}^{\Ttran}, \vect{w}_{\beta,n}^{\Ttran}]^{\Ttran}$ and measurement noise vector,  $\vect{z}_n= [\vect{z}_{\vect{r},n}^{\Ttran},\vect{z}_{\tau,n}^{\Ttran}, \vect{z}_{\mu,n}^{\Ttran}]^{\Ttran}$ with
\begin{align}
&\vect{w}_{\theta,n} = \begin{bmatrix} w_{\theta,1,n}, \ \cdots, \ w_{\theta,L,n}\end{bmatrix}^{\Ttran} \\
& \vect{w}_{d,n} = \begin{bmatrix} w_{d,1,n}, \ \cdots, \ w_{d,L,n}\end{bmatrix}^{\Ttran} \\
& \vect{w}_{\beta,n} = \big[ w_{\beta,1,1,n}, \ \cdots, \ w_{\beta,1,L,n}, \ \cdots, \nonumber\\
&\quad \ w_{\beta,L,1,n}, \ \cdots, \ w_{\beta,L,L,n}\big]^{\Ttran} \\
& \vect{z}_{\vect{r},n} = \big[ \vect{z}_{\vect{r},1,1,n}^{\Ttran}, \ \cdots, \ \vect{z}_{\vect{r},1,L,n}^{\Ttran}, \ \cdots, \nonumber\\
&\quad \vect{z}_{\vect{r},L,1,n}^{\Ttran}, \ \cdots, \ \vect{z}_{\vect{r},L,L,n}^{\Ttran}  \big]^{\Ttran} \\
& \vect{z}_{\tau,n} = \big[ z_{\tau,1,1,n}, \ \cdots, \ z_{\tau,1,L,n}, \ \cdots, \nonumber\\
&\quad z_{\tau,L,1,n}, \ \cdots, \ z_{\tau,L,L,n} \big]^{\Ttran} \\
& \vect{z}_{\mu,n} = \big[ z_{\mu,1,1,n}, \ \cdots, \ z_{\mu,1,L,n}, \ \cdots, \nonumber\\
&\quad z_{\mu,L,1,n}, \ \cdots, \ z_{\mu,L,L,n} \big]^{\Ttran}.
\end{align}

As mentioned in the system model, both $\vect{w}_n$ and $\vect{z}_n$ are zero-mean Gaussian distributed, with covariance matrices being expressed as
\begin{align}
     \vect{Q}_s &= \diag\big(\sigma_{v}^2, \sigma_{\theta,1}^2, \cdots, \sigma_{\theta_L}^2,  \sigma_{d,1}^2, \cdots,\sigma_{d,L}^2, \nonumber\\
&\quad\sigma_{\beta,1,1}^2, \cdots, \sigma_{\beta,1,L}^2,\cdots, \sigma_{\beta,L,1}^2, \cdots, \sigma_{\beta,L,L}^2  \big) \\
     \vect{Q}_{m,n} & = \diag\big(\sigma_1^2\vect{1}_{N_rL^2}, \sigma_{\tau,1,1,n}^2, \cdots, \sigma_{\tau,1,L,n}^2, \cdots, \nonumber\\
     &\quad \sigma_{\tau,L,1,n}^2, \cdots, \sigma_{\tau,L,L,n}^2,  \sigma_{\mu,1,1,n}^2, \cdots, \sigma_{\mu,1,L,n}^2, \nonumber \\&\quad \sigma_{\mu,L,1,n}^2, \cdots, \sigma_{\mu,L,L,n}^2\big).
\end{align}

In order to linearize the system models, the Jacobian matrices for both $\vect{g}(\vect{x}_{n-1})$ and $\vect{h}(x)$ need to be computed. In the following, the epoch index $n$ is omitted for simplicity. We have
\begin{align}
 \vect{G}= & \frac{\partial \vect{g}}{\partial \vect{x}}  = \begin{bmatrix} 1 & \vect{0}_{1\times L} & \vect{0}_{1\times L} & \vect{0}_{1\times L^2} \\
\frac{\partial \boldsymbol{\theta}}{\partial v} & \frac{\partial \boldsymbol{\theta}}{\partial \boldsymbol{\theta}} &    \frac{\partial \boldsymbol{\theta}}{\partial \boldsymbol{d}} & \vect{0}_{L\times L^2} \\ \frac{\partial \boldsymbol{d}}{\partial v} & \frac{\partial \boldsymbol{d}}{\partial \boldsymbol{\theta}} &   \vect{I}_{L}& \vect{0}_{L\times L^2}\\
\frac{\partial \boldsymbol{\beta}}{\partial v} & \frac{\partial \boldsymbol{\beta}}{\partial \boldsymbol{\theta}} &    \frac{\partial \boldsymbol{\beta}}{\partial \boldsymbol{d}} & \frac{\partial \boldsymbol{\beta}}{\partial \boldsymbol{\beta}} \end{bmatrix},
\end{align}
where the terms of this matrix are derived by taking partial derivatives as follows:
\begin{align}
   & \frac{\partial \boldsymbol{\theta}}{\partial v} = \begin{bmatrix} d_{1}^{-1}\Delta T \sin(\theta_1) \\
    \vdots \\
    d_{L}^{-1} \Delta T \sin(\theta_L)\end{bmatrix},  \quad \frac{\partial \boldsymbol{d}}{\partial v} = \begin{bmatrix} - 
 \Delta T \cos(\theta_1) \\
    \vdots \\
    - \Delta T \cos(\theta_L) \end{bmatrix}, \\
    & \frac{\partial \boldsymbol{\beta}}{\partial v} = \begin{bmatrix} \beta_{1,1}\left(d_1^{-1}\Delta T \cos(\theta_1) + d_1^{-1}\Delta T \cos(\theta_1)\right) \\
    \vdots \\
  \beta_{1,L}\left(d_1^{-1}\Delta T \cos(\theta_1) + d_L^{-1}\Delta T \cos(\theta_L)\right) \\
  \beta_{L,1}\left(d_L^{-1}\Delta T \cos(\theta_L) + d_1^{-1}\Delta T \cos(\theta_1)\right) \\
  \vdots \\
  \beta_{L,L}\left(d_L^{-1}\Delta T \cos(\theta_L) + d_L^{-1}\Delta T \cos(\theta_L)\right) 
  \end{bmatrix}  \label{eq:derivatives_first}\\
  & \frac{\partial \boldsymbol{\theta}}{\partial \boldsymbol{\theta}} = \diag\Bigg(  1+d_1^{-1}v\Delta T \cos(\theta_1), \cdots, \nonumber\\
  &\hspace{20mm} 1+d_L^{-1}v\Delta T \cos(\theta_L)  \Bigg), \\
  &  \frac{\partial \boldsymbol{d}}{\partial \boldsymbol{\theta}} = \diag\left( v\Delta T \sin(\theta_1), \cdots, v\Delta T \sin(\theta_L)  \right) \\
  & \frac{\partial \boldsymbol{\theta}}{\partial \boldsymbol{d}} = \diag\left(-d_1^{-2}v\Delta T \sin(\theta_1), \cdots, -d_L^{-2}v\Delta T \sin(\theta_L)\right).
\end{align}
The remaining terms are expressed using Hadamard and Kronecker products as shown in \eqref{eq:derivatives_last1}-\eqref{eq:derivatives_last3} at the top of the next page.

\begin{figure*}
\begin{align}
  & \frac{\partial \boldsymbol{\beta}}{\partial \boldsymbol{\theta}} =  \left(\vect{1}_L^{\Ttran}\kron \boldsymbol{\beta}\right) \odot \left(\diag\left(-d_1^{-1}v\Delta T  \sin(\theta_1), \cdots, -d_L^{-1}v\Delta T  \sin(\theta_L)\right)\kron \vect{1}_L\right) \nonumber \\
  &\quad \quad \quad+ \left(\vect{1}_L^{\Ttran}\kron \boldsymbol{\beta}\right) \odot \left(\vect{1}_L\kron\diag\left(-d_1^{-1}v\Delta T  \sin(\theta_1), \cdots, -d_L^{-1}v\Delta T  \sin(\theta_L)\right)\right)  \label{eq:derivatives_last1}\\ 
  & \frac{\partial \boldsymbol{\beta}}{\partial \boldsymbol{d}}= \left(\vect{1}_L^{\Ttran}\kron \boldsymbol{\beta}\right) \odot \left(\diag\left(-d_1^{-2}v\Delta T  \cos(\theta_1), \cdots, -d_L^{-2}v\Delta T  \cos(\theta_L)\right)\kron \vect{1}_L\right) \nonumber \\
  &\quad \quad \quad+ \left(\vect{1}_L^{\Ttran}\kron \boldsymbol{\beta}\right) \odot \left(\vect{1}_L\kron\diag\left(-d_1^{-2}v\Delta T  \cos(\theta_1), \cdots, -d_L^{-2}v\Delta T  \cos(\theta_L)\right)\right)  \\
  &\frac{\partial \boldsymbol{\beta}}{\partial \boldsymbol{\beta}} = \diag \bigg(1+d_1^{-1}v\Delta T \cos(\theta_1)+d_1^{-1}v\Delta T \cos(\theta_1), \cdots, 1+d_1^{-1}v\Delta T \cos(\theta_1)+d_L^{-1}v\Delta T \cos(\theta_L), \nonumber\\
  & \quad \quad \quad 1+d_L^{-1}v\Delta T \cos(\theta_L)+d_1^{-1}v\Delta T \cos(\theta_1),\cdots,  1+d_L^{-1}v\Delta T \cos(\theta_L)+d_L^{-1}v\Delta T \cos(\theta_L)  \bigg).
  \label{eq:derivatives_last3}
\end{align}
\hrulefill
\end{figure*}

We also have
\begin{align}
& \vect{H} =\frac{\partial \vect{h}}{\partial \vect{x}} = 
\begin{bmatrix}
   \vect{0}_{L^2N_r\times 1} &  \frac{\partial \Tilde{\vect{r}} }{\partial \boldsymbol{\theta}} & \vect{0}_{L^2N_r\times L} & \frac{\partial \Tilde{\vect{r}} }{\partial \boldsymbol{\beta}} \\
   \vect{0}_{L^2\times 1} & \vect{0}_{L^2\times L} & \frac{\partial \boldsymbol{\tau} }{\partial \boldsymbol{d}} & \vect{0}_{L^2\times L^2} \\
  \frac{\partial \boldsymbol{\mu} }{\partial v} & \frac{\partial \boldsymbol{\mu} }{\partial \boldsymbol{\theta}} & \vect{0}_{L^2\times L} & \vect{0}_{L^2\times L^2} 
    \end{bmatrix}
\end{align}

\begin{figure*}
\begin{align}
& \frac{\partial \Tilde{\vect{r}} }{\partial \boldsymbol{\theta}}= \sqrt{p_nG/N_t}\left(\vect{1}_L^{\Ttran}\kron \left(\boldsymbol{\beta}\kron\vect{1}_{N_r}\right)\right) \odot \vect{A}+\sqrt{p_nG/N_t}\left(\vect{1}_L^{\Ttran}\kron \left(\boldsymbol{\beta}\kron\vect{1}_{N_r}\right)\right) \odot \vect{B} \label{eq:derivatives1} \\
&\frac{\partial \Tilde{\vect{r}} }{\partial \boldsymbol{\beta}} = \sqrt{p_nG/N_t}\mathrm{bdiag}\bigg(\vect{b}(\theta_{1})\vect{a}^{\Htran}(\theta_1)\vect{a}(\hat{\theta}_1),\cdots, \vect{b}(\theta_{1})\vect{a}^{\Htran}(\theta_L)\vect{a}(\hat{\theta}_L), \vect{b}(\theta_{L})\vect{a}^{\Htran}(\theta_1)\vect{a}(\hat{\theta}_1), \cdots, \vect{b}(\theta_{L})\vect{a}^{\Htran}(\theta_L)\vect{a}(\hat{\theta}_L) \bigg) \label{eq:derivatives2} \\
&\frac{\partial \boldsymbol{\mu} }{\partial v} = \frac{f_c}{c}\begin{bmatrix} (\cos(\theta_1)+\cos(\theta_1)), \cdots, (\cos(\theta_1)+\cos(\theta_L)), (\cos(\theta_L)+\cos(\theta_1)),\cdots, (\cos(\theta_L)+\cos(\theta_L))   \end{bmatrix}^{\Ttran}. \label{eq:derivatives4}
\end{align}
\hrulefill
\end{figure*}

\noindent where the partial derivatives are derived and the terms are expressed as 
\begin{align}
&\frac{\partial \boldsymbol{\tau}}{\partial \boldsymbol{d}} = \frac{1}{c} \vect{I}_L\kron \vect{1}_L +\frac{1}{c} \vect{1}_L\kron \vect{I}_L, \\
&\frac{\partial \boldsymbol{\mu} }{\partial \boldsymbol{\theta}} =  \frac{vf_c}{c} \vect{C}\kron \vect{1}_L +\frac{vf_c}{c} \vect{1}_L\kron \vect{C}, \label{eq:derivatives3} 
\end{align}
and as shown in \eqref{eq:derivatives1}-\eqref{eq:derivatives4} with
\begin{align}
  &\dot{\vect{b}}(\theta) = \frac{\partial \vect{b}}{\partial \theta}= \begin{bmatrix}
   0 \\
   \imagunit \pi \sin(\theta)e^{-\imagunit \pi\cos(\theta)} \\ 
   \vdots \\ 
 \imagunit \pi(N_r-1) \sin(\theta)e^{-\imagunit \pi(N_r-1)\cos(\theta)}  \\ 
    \end{bmatrix},\\
    &\overline{\vect{a}}=\begin{bmatrix} \vect{a}^{\Htran}(\theta_1)\vect{a}(\hat{\theta}_1), \ \cdots, \ \vect{a}^{\Htran}(\theta_L)\vect{a}(\hat{\theta}_L)  \end{bmatrix}^{\Ttran} \\
    &\dot{\overline{\vect{a}}}=\begin{bmatrix}
    -\sum\limits_{i=1}^{{N}_t} e^{-\imagunit \pi(i-1)\cos{\hat{\theta}_{1}}+\imagunit \pi(i-1)\cos{\theta_1}}\imagunit\pi(i-1)\sin(\theta_1)   \\
   \vdots \\
   -\sum\limits_{i=1}^{{N}_t} e^{-\imagunit \pi(i-1)\cos{\hat{\theta}_{L}}+\imagunit\pi(i-1)\cos{\theta_L}}\imagunit\pi(i-1)\sin(\theta_L)  
    \end{bmatrix} \\
    & \vect{A} = \mathrm{bdiag}\left( \overline{\vect{a}}\kron \dot{\vect{b}}(\theta_1),\cdots, \overline{\vect{a}}\kron \dot{\vect{b}}(\theta_L)\right)  \\
    & \vect{B} = \mathrm{bdiag}\left( \dot{\overline{\vect{a}}}\kron \vect{b}(\theta_1),\cdots, \dot{\overline{\vect{a}}}\kron \vect{b}(\theta_L)\right) \\
    &\vect{C}=\mathrm{diag}(-\sin(\theta_1), \ldots,-\sin(\theta_L)).
\end{align}

After the Jacobian matrices are computed, the EKF technique can be applied easily. Although the following steps are conventional for EKF filtering, designed state evaluation and measurement models are implemented as shown in Table \ref{tab:EKF}.\begin{table*}[h]
  \centering  \caption{Extended Kalman Filtering Procedure}
    \label{tab:EKF}
    \centering
    \begin{tabular}{|c||c|}
    \hline
    \textbf{Step} &  \\
    \hline
     State Prediction & $\hat{\vect{x}}_{n|n-1} = \vect{g}(\hat{\vect{x}}_{n-1})$, $\hat{\vect{x}}_{n|n-2} = \vect{g}(\hat{\vect{x}}_{n-1|n-2})$  $\hat{\vect{x}}_{n-1|n-2} = \vect{g}(\hat{\vect{x}}_{n-2})$\\ 
     \hline
     Linearization & $  
     \vect{G}_{n-1} = \frac{\partial{\vect{g}}}{\partial{x}}
     |_{\vect{x}=\hat{\vect{x}}_{n-1}}
    , 
      
     \vect{H}_{n} = \frac{\partial{\vect{h}}}{\partial{x}}|_{\vect{x}=\hat{\vect{x}}_{n|n-1}}
      $ 
     \\
     \hline
     MSE Matrix Prediction 
     & $ \vect{M}_{n|n-1} = \vect{G}_{n-1} \vect{M}_{n-1} \vect{G}_{n-1}^{\Htran} + \vect{Q}_s $  
      \\ 
     \hline
     Kalman Gain Calculation & $ \vect{K}_{n} = \vect{M}_{n|n-1} \vect{H}_{n}^{\Htran} (\vect{Q}_m + \vect{H}_{n} \vect{M}_{n|n-1} \vect{H}_{n}^{\Htran} )^{-1}$ \\ 
     \hline
     State Tracking & $\hat{\vect{x}}_{n} = \hat{\vect{x}}_{n|n-1} + \vect{K}_{n} (\vect{y}_n - \vect{h}(\hat{\vect{x}}_{n-1}))$  \\ 
     \hline
     MSE Matrix Update & $ \vect{M}_{n} = (\vect{I} -  \vect{K}_{n}\vect{H}_{n})\vect{M}_{n|n-1} $  \\ 
     \hline
    \end{tabular}

\end{table*}

In this work, we assume that all $L$ RSUs contribute to radar sensing and transmission. In reality, the vehicle would be traveling along a long road with many potential RSUs along the way. In that case, it would be reasonable to select the closest or most suitable set of RSUs and update this set during the course of travel. However, in this work, we aim to formulate the Kalman filter equations and observe the advantage of predictive beamforming with distributed MIMO. RSU selection will be the subject of future work. 

\section{Numerical Results} \label{Numerical Results}

In this section, we will quantify the performance of the proposed predictive beamforming scheme with distributed MIMO compared to the benchmark scheme, where the latter is the case of a single MIMO transceiver (RSU). Unless otherwise specified, default system parameters are shown in Table \ref{tab:simparams}. Following \cite{Liu2020RadarAssisted}, we set the measurement noise parameters as  $\sigma_1^2=\sigma_c^2/(p_nG) \, \text{W}$, $\sigma_2^2=6.7\cdot 10^{-7} \text{sec}^2\cdot\text{W}$, and $\sigma_3^2=2\cdot10^4 \text{Hz}^2\cdot\text{W}$. We assume a 3GPP UMi Street Canyon pathloss model.  We make $1000$ trials, and at each of them, the horizontal RSU coordinates are regenerated randomly according to the uniform distribution given in Table~\ref{tab:simparams}. At each trial, the vehicle starts at a certain coordinate and moves left with a fixed velocity. We plot i) the cumulative distribution function (CDF) of the SE (in bps/Hz) and ii) the average SE vs. time. We do this for different antenna configurations. For example, $L=4, N_t=4, N_r=4$ means there are $4$ RSUs with $4$ transmit and receive antennas at each. Note that $L=4, N_t=4, N_r=4$ and $L=1, N_t=16, N_r=16$ have the same total number of antennas for a fair comparison between co-located predictive beamforming and distributed predictive beamforming.

\begin{figure*}[h]
    \centering
    \includegraphics[width=2\columnwidth]{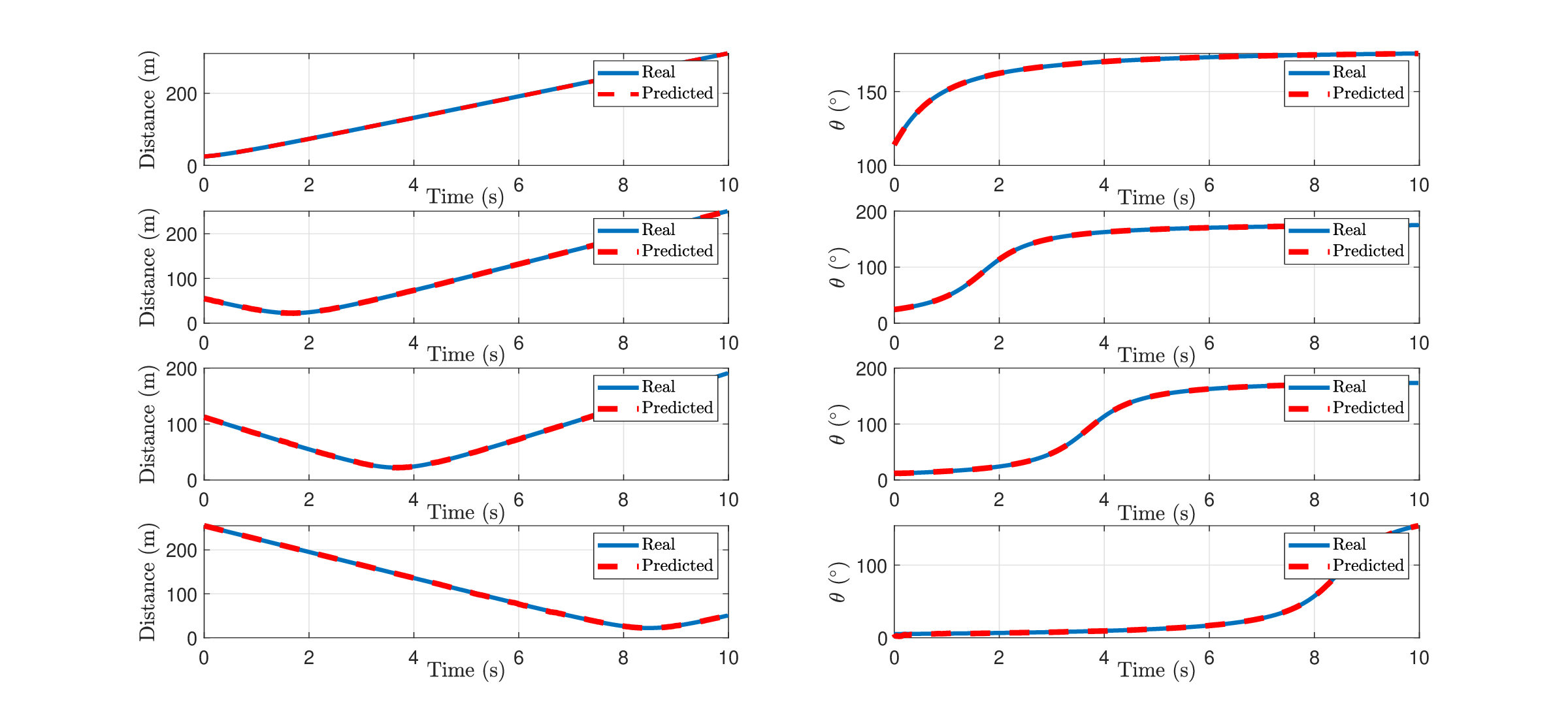}
    \caption{Actual and predicted parameters by $L=4$ RSUs with $N_t=N_r=16$ antennas, while the vehicle is moving from right to left. Rows correspond to each RSU, while columns correspond to distance and angle.} 
    \label{Parameters}
\end{figure*}

Fig.~\ref{Parameters} illustrates the angle and distance tracking performance. Using the state evaluation model, the predicted distance and angle values closely match the actual values. Other parameters of the state evaluation model are not presented in this paper because the velocity is constant, and the RCS depends on both distance and angle. If these two parameters are accurately predicted, the RCS will also be correctly estimated. Additionally, it is worth noting that the parameters of each RSU differ due to their relative positions to the vehicle. 
\begin{table}[htb]
  \caption{Simulation Parameters}
    \label{tab:simparams}
    \centering
    \begin{tabular}{|c|c|}
    \hline
      \textbf{Parameter}   & \textbf{Value}  \\\hline
      Number of RSU antennas  ($N_t=N_r$)  &  4, 16, 64 \\\hline
      Number of RSUs    &  $1$ or $4$ \\\hline
      Number of vehicle antennas  ($M$)  &  $8$ \\\hline
      RSU position $(x,y)$   &  $\begin{bmatrix}(\mathcal{U}(0,35), -22.5)\\(\mathcal{U}(50,85), 22.5)\\(\mathcal{U}(135,170), -22.5)\\(\mathcal{U}(270,300), 22.5)\end{bmatrix}$ \\\hline
      Vehicle start position $(x,y)$   &  $(35,0)$ \\\hline
      Carrier frequency ($f_c$)  & $30$ GHz \\\hline
      Number of time slots ($N$)   & $500$  \\\hline
      Time slot duration ($\Delta T$)  & $0.02$ sec \\\hline
      Velocity ($v$)    & $30$ m/sec  \\\hline
      $\sigma_{\theta,l}$  & $[0.02, 0.015, 0.022, 0.018]^{\circ}$ \\\hline
      $\sigma_{d,l}$     & $[0.2, 0.18, 0.21, 0.19]$ \\\hline
      $\sigma_v$     & $0.5$\\\hline
      $\sigma_{\beta,l,m}$     & $\begin{bmatrix}
                            0.1 &  0.09 & 0.11 & 0.1 \\
                            0.12 & 0.08 & 0.13 & 0.08 \\
                            0.11 & 0.09 & 0.1 & 0.1 \\
                            0.08 & 0.09 & 0.12 & 0.09\end{bmatrix}$ \\\hline
      Noise variance ($\sigma_c^2$)     & $-75$\,dBm \\\hline
      Matched filtering gain ($G$) & $10$\\\hline
      Transmit power ($p_n$)          & $1$\,W \\\hline
    \end{tabular}
  
\end{table}

\begin{figure}[h]
    \centering
    \includegraphics[width=3.5in]{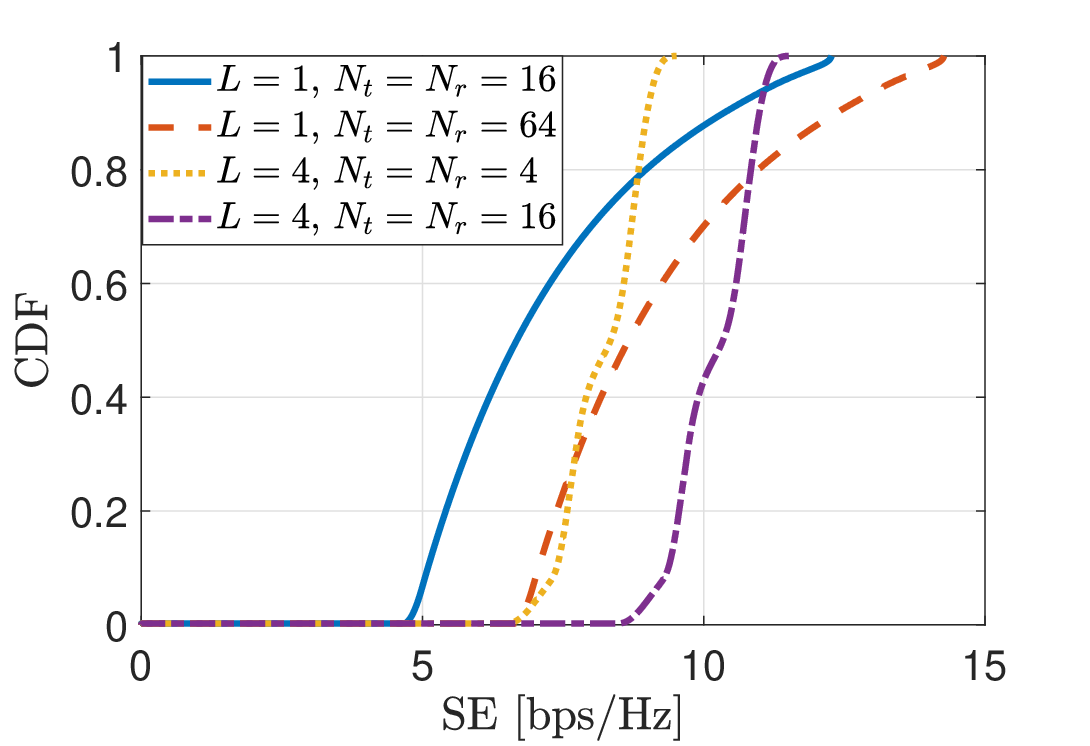}
    \caption{The CDF of the SE for different configurations.} 
    \label{fig:CDFvsSE}
\end{figure}

Fig. \ref{fig:CDFvsSE} presents the CDF of SE for various antenna configurations. The results demonstrate that the $L=4, N_t=4, N_r=4$ configuration (i.e., distributed MIMO) achieves a higher median SE compared to the $L=1, N_t=16, N_r=16$ configuration (i.e., a single RSU with a larger number of antennas). This improvement is due to the enhanced geographical coverage provided by the distributed antennas, resulting in superior sensing and communication performance. Similarly, the $L=4, N_t=16, N_r=16$ configuration outperforms its co-located single RSU counterpart in both 90\%-likely and median SE. Hence, sensing-assisted distributed MIMO communication is more advantageous for providing a more uniform data rate.

\begin{figure}[h]
    \centering
    \includegraphics[width=3.5in]{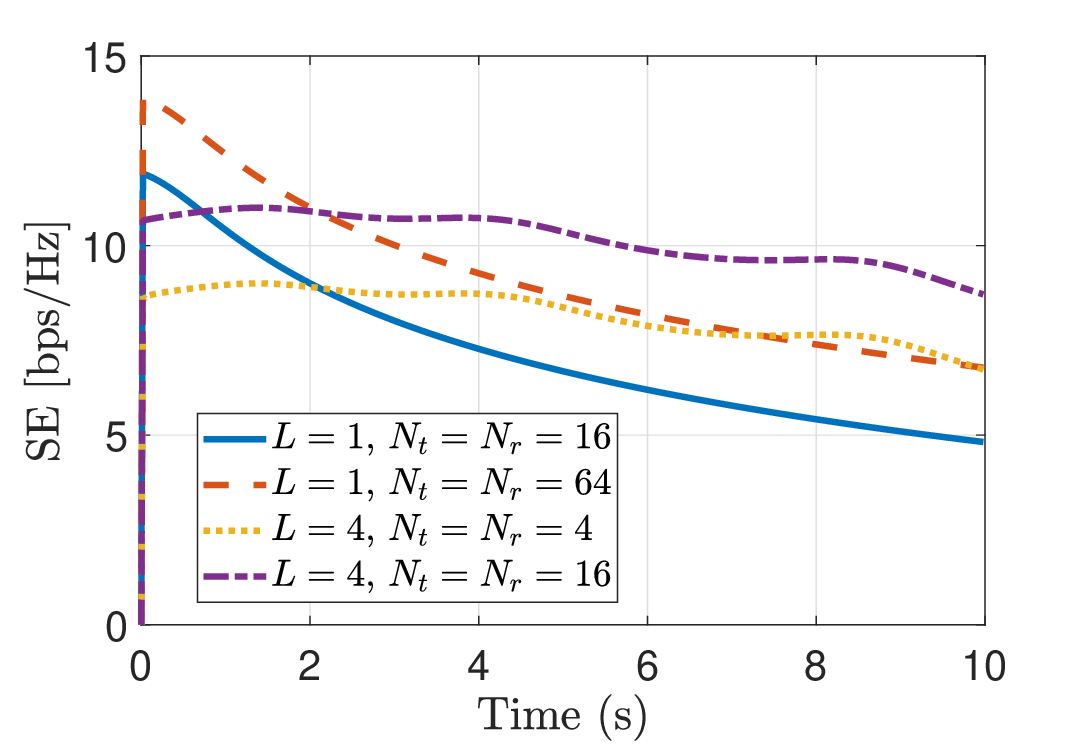}
    \caption{The SE versus time as the vehicle moves.} 
    \label{fig:SEvsTime}

\end{figure}

Fig.  \ref{fig:SEvsTime}  illustrates the average SE over time as the vehicle moves. The $L=1$ configuration with a large number of antennas achieves better SE when the vehicle is nearby. However, distributed MIMO provides more uniform communication performance thanks to the more robust sensing functionality. The smaller variations in data rate are advantageous for simplifying modulation and coding schemes.

\section{Conclusions} \label{Conclusions}

This paper presents the first implementation of sensing-assisted predictive beamforming in the context of distributed MIMO. We derived the state evolution and Kalman filter equations and compared the SE of distributed MIMO with that of a single RSU with the same total number of antennas. The results demonstrate that sensing-assisted predictive beamforming is effective in a distributed MIMO setting. Furthermore, distributed MIMO can achieve more uniform SE as the vehicle moves without compromising sensing performance.

Several directions for future work are worth exploring. In this study, we assumed that each RSU transmits in orthogonal channels to avoid excessive complexity in derivations and implementations. However, using joint transmission could potentially achieve even greater SE. Additionally, real-world scenarios may involve momentary blockages in mmWave, which could significantly impact radar and communication performance. Distributed MIMO is expected to offer extra robustness in such situations.

\bibliographystyle{IEEEtran}
\bibliography{IEEEabrv,refs}

\end{document}